\documentclass[twocolumn,showpacs,showkeys,amsmath,amssymb,superscriptaddress,floatfix,nofootinbib]{revtex4}

\usepackage{graphicx}
\usepackage{bm}
\usepackage{subfigure}

\makeatletter

\begin{document}
 
\title{
Size fluctuations of the initial source and the event-by-event 
transverse momentum fluctuations in relativistic heavy-ion collisions%
\footnote{Supported in part by the Polish Ministry of Science and Higher
Education, grants N202~034~32/0918 and N202~249235.}}

\author{Wojciech Broniowski} 
\email{Wojciech.Broniowski@ifj.edu.pl} 
\affiliation{The H. Niewodnicza\'nski Institute of Nuclear Physics, Polish Academy of Sciences, PL-31342 Krak\'ow, Poland}
\affiliation{Institute of Physics, Jan Kochanowski University, PL-25406~Kielce, Poland} 

\author{Mikolaj Chojnacki}
\email{Mikolaj.Chojnacki@ifj.edu.pl}
\affiliation{The H. Niewodnicza\'nski Institute of Nuclear Physics, Polish Academy of Sciences, PL-31342 Krak\'ow, Poland}

\author{\L{}ukasz Obara} 
\email{lukaszo4@onet.eu}
\affiliation{Institute of Physics, Jan Kochanowski University, PL-25406~Kielce, Poland} 

\date{ver. 2, 1 October 2009}

\begin{abstract}
We show that the event-by-event fluctuations of the transverse size of the initial source, which follow 
directly from the Glauber treatment 
of the earliest stage of relativistic heavy-ion collisions, cause, after hydrodynamic evolution, fluctuations of the 
transverse flow velocity at hadronic freeze-out. This in turn leads to event-by-event fluctuations of the average transverse momentum, 
$\langle p_T \rangle$. Simulations 
with {\tt GLISSANDO} for the Glauber phase, followed by a realistic hydrodynamic evolution and statistical hadronization 
carried with {\tt THERMINATOR}, lead to agreement with the RHIC data. In particular, the magnitude of the effect, 
its centrality dependence, and the weak dependence on the incident energy are properly reproduced. Our results show that 
bulk of the observed event-by-event $\langle p_T \rangle$ fluctuations may be explained by the fluctuations of the size of the initial source.
\end{abstract}

\pacs{25.75.-q, 25.75.Dw, 25.75.Ld}

\keywords{relativistic heavy-ion collisions, transverse-momentum fluctuations, Glauber models, hydrodynamics, 
statistical hadronization, SPS, RHIC, LHC}

\maketitle 

We propose a new mechanism for generating the transverse-momentum fluctuations in relativistic heavy-ion collisions, based 
on the fluctuations of the initial size of the formed system and its subsequent hydrodynamic evolution.
It is well established that a successful description of the physics of relativistic heavy-ion collisions is 
achieved with the help of relativistic hydrodynamics \cite{Kolb:2000sd,Bass:2000ib,Teaney:2000cw,Kolb:2001qz,Hirano:2002ds,Huovinen:2003fa,Kolb:2003dz,Shuryak:2004cy,%
Muller:2007rs,Nonaka:2007nn,Broniowski:2008vp,Pratt:2008bc}, effective at 
proper times ranging typically from about a fraction of a fm/c to a 
few fm/c, 
where statistical hadronization takes over. 
Numerous observables can be reproduced that way, such as 
the momentum spectra, elliptic flow, or the HBT correlation radii \cite{Broniowski:2008vp,Pratt:2008bc}, measuring the system's space-time
extension. 
The initial condition for hydrodynamics is usually obtained 
from the Glauber approach, leading to the wounded-nucleon picture \cite{Bialas:1976ed} 
(a wounded nucleon is a nucleon, that has collided inelastically 
at least once) or its variants \cite{Kharzeev:2000ph,Broniowski:2007nz}. When the initial 
condition is obtained via Glauber Monte Carlo simulations, its shape {\em fluctuates}, 
simply reflecting the randomness in positions of the 
nucleons in the colliding nuclei. 

In this paper we show that the event-by-event 
fluctuations of the initial size are substantial, even when we consider the class of events with a strictly fixed 
number of the wounded nucleons, $N_w$.
The fluctuations are then carried over by hydrodynamics to the fluctuations of the transverse flow velocity
at the hadronic freeze-out, which in turn generate the event-by-event fluctuations of the average transverse momentum, $p_T$, of the 
produced hadrons. The mechanism is very simple: a more {\em squeezed} initial condition leads to faster expansion, larger flow, and, 
consequently, higher $\langle p_T \rangle$, while, on the conrary, a more {\em stretched} initial condition leads to slower expansion, 
lower flow, and lower $\langle p_T \rangle$. 
The event-by-event $p_T$ fluctuations have been a subject of intense theoretical and experimental studies
\cite{Gazdzicki:1992ri,Stodolsky:1995ds,Shuryak:1997yj,%
Mrowczynski:1997kz,%
Voloshin:1999yf,Baym:1999up,Voloshin:2001ei,Prindle:2006zz,Mrowczynski:2009wk,%
Adams:2003uw,Adamova:2003pz,Adler:2003xq,Adams:2005ka,Grebieszkow:2007xz,na49:2008vb}, as they may reveal 
important details of the dynamics of the system.
Effects of inhomogeneities for various observables have been studied in Ref.~\cite{Hama:2009pk}.
The event-by-event fluctuations of the initial shape have been studied in detail for its elliptic
component, where they lead to enhanced elliptic flow \cite{Aguiar:2000hw,Miller:2003kd,Bhalerao:2005mm,Andrade:2006yh,Voloshin:2006gz,Broniowski:2007ft,Hama:2007dq,Voloshin:2007pc,%
Manly:2005zy,Alver:2006wh}. Our present study is similar in spirit, but it 
focuses on the size fluctuations.

\begin{figure}[tb]
\begin{center}
\includegraphics[angle=0,width=0.33 \textwidth]{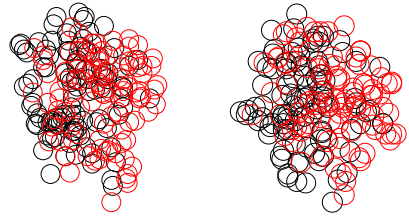}
\end{center}
\vspace{-6mm}
\caption{(Color online) Two sample collisions, both with equal number of wounded nucleons, $N_w=198$, displayed as circles. 
The left and right cases have 
$\langle r \rangle=2.95$~fm and $2.83$~fm, respectively. \label{fig:sample}} 
\end{figure}

Consider the average transverse size, defined in each event as (we use the
wounded nucleon model for the simplicity of notation)
\begin{eqnarray}
\langle r \rangle = \sum_{i=1}^{N_w} \sqrt{x_i^2+y_i^2}, \label{eq:def}
\end{eqnarray}  
where $x_i$ and $y_i$ are coordinates of a wounded nucleon in the transverse plane. Examples of the spatial distributions 
of wounded nucleons are shown if Fig.~\ref{fig:sample}, 
where collisions of two ${}^{197}$Au nuclei are viewed along the beam. 
We note that the two cases displayed in Fig.~\ref{fig:sample}, although having equal numbers of wounded nucleons, 
have indeed a rather different shape and size.     
The original positions of nucleons in each 
nucleus are randomly generated from an appropriate Woods-Saxon distribution, with an additional constraint that the centers of nucleons
in each nucleus 
cannot be closer than the expulsion radius $d=0.4$~fm, which simulates the short-range repulsion.
Nucleons from the two nuclei get wounded or undergo a binary collision 
when their centers pass closer to each other than the distance 
$d=\sqrt{\sigma_{NN}/\pi}$, where $\sigma_{NN}$ is the inelastic nucleon-nucleon cross section. For the highest 
SPS, RHIC, and LHC energies it is equal to 32, 42, and 63~mb, respectively. 

\begin{figure}[tb]
\begin{center}
\includegraphics[angle=0,width=0.47 \textwidth]{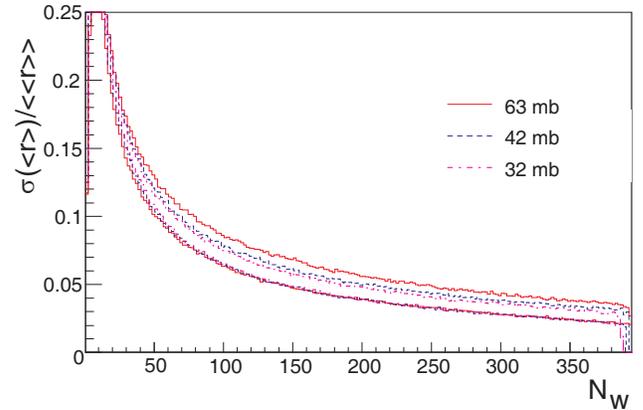} 
\end{center}
\vspace{-3.5mm}
\caption{(Color online) 
Event-by-event scaled standard deviation of the size parameter $\langle r \rangle$, evaluated at fixed values 
of the number of wounded nucleons, $N_w$. The three lower overlapping curves are for the wounded nucleon model at 
the NN cross section corresponding to the highest SPS (32~mb), RHIC (42~mb), and LHC (63~mb) energies. The three upper curves are for the 
mixed model at the subsequent energies. The dependence on the energy for the mixed model originates entirely from different values of the $\alpha$ 
mixing parameter, see the text for details. 
\label{fig:basic}} 
\end{figure}

We introduce the notation $\langle \langle . \rangle \rangle$ to indicate averaging over the events. 
In order to focus on the relative size of the 
effect we use the scaled standard deviation, defined for a fixed value of $N_w$ 
as $\sigma(\langle r \rangle)/\langle \langle r \rangle \rangle$. The results of our Monte Carlo simulations
performed with {\tt GLISSANDO} are shown in Fig.~\ref{fig:basic}. We present the standard wounded nucleon 
model \cite{Bialas:1976ed} (the three lower overlapping curves) and the mixed model \cite{Kharzeev:2000ph,Broniowski:2007nz} 
(three upper curves), where the Glauber source distribution is 
formed from the fractions $(1-\alpha)/2$ of the wounded nucleons and $\alpha$ of the binary collisions. 
The mixing parameter $\alpha$ is assumed to be equal to 0.12, 0.145, and 0.2 for the highest SPS, RHIC, and LHC energies, respectively.
From the fact that the three curves for the wounded nucleon model overlap, we conclude that the effect is completely insensitive 
to the value of $\sigma_{NN}$ within the considered range. 
The observed moderate dependence on the energy for the mixed model originates entirely from the different values of the $\alpha$
parameter. For the binary collisions the size fluctuations are stronger than for the wounded nucleons, hence a larger value of $\alpha$ 
yield larger fluctuations.
We note that the scaled variance of $\langle r \rangle$ is 
about 2.5-3.5\% for central collisions, and grows towards the peripheral collisions approximately as $1/\sqrt{N_w}$.
Very similar results are obtained for other variants of Glauber models, in particular models with 
overlaid distribution of particles produced by each wounded nucleon \cite{Broniowski:2007nz}. We have also 
checked that using a Gaussian wounding profile $\sigma_{NN}(b)$ \cite{Bialas:2006kw} for the 
$NN$ collisions, rather than the 
sharp wounding distance criterion applied here, leads to very similar curves. Furthermore,   
the use of the nucleon distributions including realistically the central $NN$ correlations, as given in Ref.~\cite{Alvioli:2009ab}, 
leads to practically no difference. In other words, the behavior displayed in Fig.~\ref{fig:basic}
is robust, basically reflecting the statistical nature of the Glauber approach.

The next step, crucial in converting the size fluctuation into momentum fluctuations, is hydrodynamics. 
We use the perfect boost-invariant hydrodynamics described in detail in Ref.~{\cite{Broniowski:2008vp}, which 
leads to a successful simultaneous description of the
soft RHIC observables at midrapidity, such as the $p_T$ spectra, elliptic flow, and the HBT radii, including the 
azimuthally-sensitive (azHBT) femtoscopy \cite{Kisiel:2008ws}.
The essential ingredients of this
approach are the Gaussian transverse initial condition, early
start of the evolution ($\tau_0=0.25$~fm/c), and the state-of-the art equation of state \cite{Broniowski:2008vp}, incorporating a
smooth crossover and interpolating between the lattice QCD results at high temperatures and the resonance gas at low 
temperatures. The initial transverse energy-density profile is assumed to have the simple form 
%
\mbox{$\epsilon(x,y)=\epsilon_0 \exp [ -{x^2}/({2a^2}) -{y^2}/({2 b^2}) ]$},
%
where $x$ and $y$ denote the transverse coordinates.
The width parameters $a$ and $b$ depend on centrality. To obtain realistic 
values we run {\tt GLISSANDO} \cite{Broniowski:2007nz} simulations, which include the eccentricity 
fluctuations. Then we match $a$ and $b$ to reproduce the values
$\langle x^2 \rangle$ and $\langle y^2 \rangle$ from the {\tt GLISSANDO} Monte Carlo profiles. 
Thus, by construction, the spatial rms radii of the initial condition and its eccentricity are the same as 
from the Glauber calculation. 
The values of the used width parameters are given in Table~\ref{tab:ab}. The central 
value of the initial temperature is adjusted in such a way 
that the multiplicity of produced particles is reproduced. 

\begin{table}[tb]
\caption{Shape parameters $a$ and $b$, as well as the central temperature $T_i$, for various centrality classes
for the highest RHIC energy of $\sqrt{s_{NN}}=200$~GeV. 
\label{tab:ab}}
\begin{tabular}{r|rrrrrrrrr}
\hline
$c$ [\%] 		&  0- 5 &  5-10 & 10-20 & 20-30 & 30-40 & 40-50 & 50-60 & 60-70 \\
\hline
$a$ [fm]		& 2.70  & 2.54  & 2.38  & 2.00  & 1.77  & 1.58  & 1.40  & 1.22  \\
$b$ [fm] 		& 2.93  & 2.85  & 2.74  & 2.59  & 2.45  & 2.31  & 2.16  & 2.02   \\
$T_i$ [MeV]     & 500   & 491   & 476   & 455   & 429   & 398   & 354   & 279    \\
\hline
\end{tabular}
\end{table}

The average transverse momentum is also reproduced by our model, as can be seen from Fig.~\ref{fig:pt}, where we compare 
the model predictions (solid line) to the STAR data \cite{abelev:2008ez}, extrapolated to the full $p_T$ coverage. 
The dashed line corresponds to the model calculation with the STAR range $0.2~{\rm GeV} < p_T < 2~{\rm GeV}$ used in 
Ref.~\cite{Adams:2005ka}. 

\begin{figure}[b]
\begin{center}
\includegraphics[angle=0,width=0.47 \textwidth]{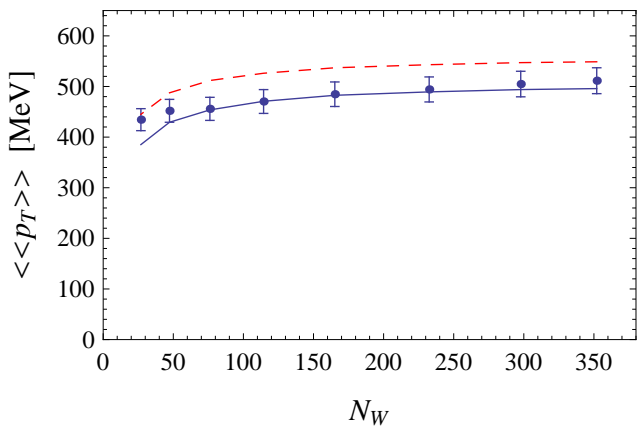} 
\end{center}
\vspace{-5mm}
\caption{(Color online) Dependence of the average transverse momentum, $\langle \langle p_T \rangle \rangle$,
on the number of wounded nucleons. The solid (dashed) line corresponds to the model calculation 
in the full $p_T$ range (in the range of the STAR coverage $0.2~{\rm GeV} < p_T < 2~{\rm GeV}$).
The experimental points (obtained by extrapolating the $p_T$ distributions to the full range) 
are taken from Ref.~\cite{abelev:2008ez}. \label{fig:pt}}
\end{figure}

Next, we analyze the hydrodynamic evolution with fluctuating initial conditions. Rather than doing an event-by-event calculation, 
which is tedious, it suffices (see the following) to see how much the predictions change when the $a$ and $b$ parameters are scaled by the values 
read off from Fig.~\ref{fig:basic}. For instance, for $c=20-30\%$, which corresponds to $N_w\simeq 165$, we note from Fig.~\ref{fig:basic}
that the scaled standard deviation of the size fluctuations is 4.4\% (for the wounded nucleon model). 
Thus we rescale $a$ and $b$ up and down by 4.4\% and run 
the simulations. In addition, we also adjust the value of the central temperature $T_i$ in such a way, that the energy contained in the
profile is preserved. This is natural, as the total energy deposited in the transverse plane should be 
(up to possible additional fluctuations) the same for a given 
number of elementary collisions. Hence a squeezed system has a higher central temperature than a stretched system. 
Thus, in some sense, we include also the temperature fluctuations discussed in Ref.~\cite{Korus:2001au}.
Additional event-by-event energy fluctuations (in the considered pseudorapidity window $|\eta|<1$) could be added on top of the 
analyzed effect, which 
would act as another source of momentum fluctuations, not included in this work. 

Hydrodynamics is run till the local temperature drops to $T_f=145$~MeV \cite{Broniowski:2008vp}, 
where freezeout occurs.  The 
freezeout hypersurfaces in the space of the transverse radius, $r$,  and time, $t$, 
at $z=0$, as well as the transverse velocities, for the 5\% 
squeezed and stretched cases for $c=20-30\%$ are shown in Fig.~\ref{fig:hs}. We note the following features: the 
maximum expansion velocities, indicated by dots with labels, are about 10\% larger for the squeezed case compared to the stretched case, in addition 
the squeezed case is somewhat more compact, {i.e.} the radius $r$ and the time $t$ at freezeout are slightly smaller.

\begin{figure}[tb]
\begin{center}
\includegraphics[angle=0,width=0.44 \textwidth]{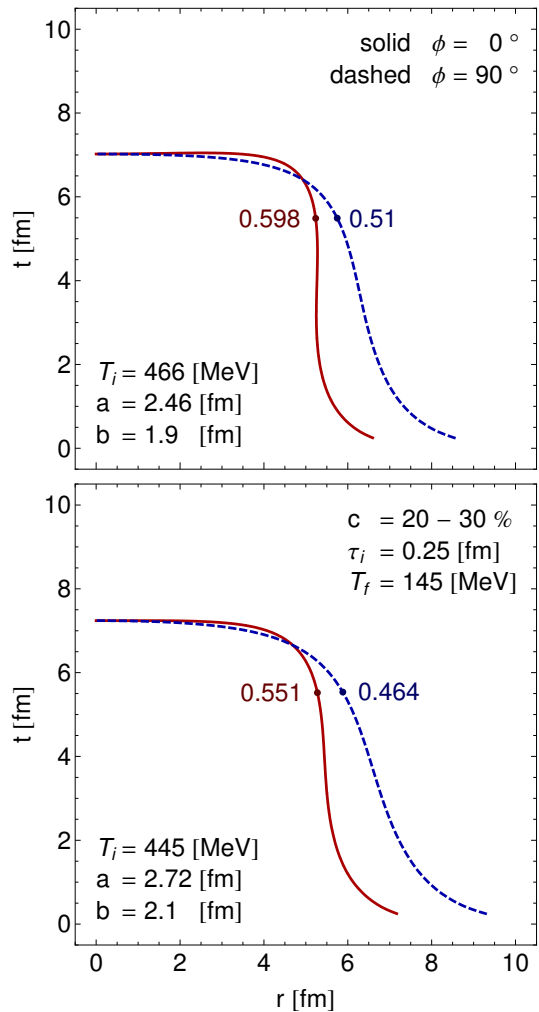} 
\end{center}
\vspace{-5.5mm}
\caption{(Color online)  
In-plane (solid) and out-of-plane (dashed) freezeout hypersurfaces in the transverse coordinate-time plane for $c=20-30\%$, with the
indicated values of the maximum transverse flow velocity in units of $c$. Top (bottom): the 5\% squeezed (stretched) 
initial condition.  
\label{fig:hs}} 
\end{figure}

The final stage of our calculation, turning the transverse-velocity fluctuations into the 
transverse-momentum fluctuations,  is the simulation of the statistical hadronization at freezeout 
with {\tt THERMINATOR} \cite{Kisiel:2005hn}, which 
includes all resonances and decay channels from {\tt SHARE} \cite{Torrieri:2004zz}.
According to the Cooper-Frye formalism, the particles (stable and unstable, which subsequently 
decay) are formed at the freezeout hypersurface according to appropriate statistical distributions. 
For the squeezed case they are more ``pushed'' with the larger flow velocity than for the stretched case, 
thus they acquire a higher average transverse momentum, $\langle \langle p_T \rangle \rangle$. 
The experimental cuts of the STAR detector are used, with $0.2~{\rm GeV} < p_T < 2~{\rm GeV}$.

In order to compare to the data, we analyze the STAR correlation measure \cite{Adams:2005ka}, 
$\sigma_{\rm dyn}^2 \equiv \langle \Delta p_i \Delta p_j \rangle$. It can be shown to be equal to 
\begin{eqnarray}
\sigma^2_{\rm dyn}=\sigma^2(\langle p_T \rangle)-\frac{1}{N_{\rm ev}}\sum_{k=1}^{N_{\rm ev}} \frac{\sigma^2_k(p_T)}{N_k},
\end{eqnarray}
where $k$ is numbers the events, $\sigma^2_k(p_T)$ the variance of $p_T$ in a given event, and $N_k$ denotes the multiplicity of the event.
By construction, 
the second term cancels the uncorrelated (purely statistical) fluctuations in the first term, leaving in $\sigma_{\rm dyn}$ the dynamical 
correlations only. Our simulations do not include the statistical event-by-event 
fluctuations of $\langle p_T \rangle$, which would follow from the random nature of the statistical hadronization involving
a finite number of particles. The procedure described below 
allows to avoid the tedious event-by-event studies in the extraction of $\sigma_{\rm dyn}$ 
when it originates form the fluctuations of the initial condition. 

The full statistical distribution 
$f(\langle p_T \rangle)$ in a given centrality class is a folding of the statistical distribution of $\langle p_T \rangle$ at a fixed 
initial size, centered around a certain $\bar p_T$, with the distribution of $\bar p_T$ centered around $\langle\langle p_T \rangle\rangle$. 
The value of $\bar p_T$, corresponding to a fixed initial size, fluctuates because of the fluctuations of the initial size.  
In the central 
regions both are close to gaussian distributions, hence we have to a very good approximation
\begin{eqnarray}
&&f(\langle p_T \rangle) \sim \int d^2{\bar{p}_T} \times \label{fp} \\ && 
\exp{\left ( -\frac{(\langle p_T \rangle - {\bar p}_T)^2}{2 \sigma_{\rm stat}^2} \right )}
\exp{\left ( -\frac{({\bar p}_T  - \langle \langle p_T \rangle \rangle)^2}{2 \sigma_{\rm dyn}^2}\right ) }. \nonumber
\end{eqnarray}
Carrying out the $\bar{p}_T$ integration yields the distribution of $\langle p_T \rangle$ centered around $\langle \langle p_T \rangle \rangle$
with the width parameter satisfying  $\sigma^2=\sigma_{\rm stat}^2+ \sigma_{\rm dyn}^2$. With the above factorized form 
we may obtain the second term directly from the distribution of the initial size parameter $\langle r \rangle$. Its distribution is also approximately gaussian, 
\begin{eqnarray}
f(\langle r \rangle) \sim \exp \left (-\frac{(\langle r \rangle- \langle \langle r \rangle \rangle)^2}{2 \sigma^2(\langle r \rangle)} \right ).
\label{fr}
\end{eqnarray} 
Because of the deterministic nature of hydrodynamics, $\bar p_T$ is a (complicated) function of $\langle r \rangle$. 
Nevertheless, in the vicinity of the central values 
we have from the Taylor expansion
\begin{eqnarray}
\bar p_T - \langle \langle p_T \rangle \rangle = 
\left . \frac{d \bar p_T}{d\langle r \rangle} \right |_{\langle r \rangle = \langle \langle r \rangle \rangle} 
\left ( \langle r \rangle - \langle \langle r \rangle \rangle\right ) +\dots \label{fexp}
\end{eqnarray} 
Substituting (\ref{fexp}) into (\ref{fr}) and comparing to (\ref{fp}) we obtain the key result
\begin{eqnarray}
\sigma_{\rm dyn} = \sigma(\langle r \rangle) 
\left . \frac{d \bar p_T}{d\langle r \rangle} \right |_{\langle r \rangle = \langle \langle r \rangle \rangle},
\label{sigsig}
\end{eqnarray}
or for the scaled standard deviation
\begin{eqnarray}
\frac{\sigma_{\rm dyn}}{\langle \langle p_T \rangle \rangle} = \frac{\sigma(\langle r \rangle)}{\langle \langle r \rangle \rangle}
\frac{\langle \langle r \rangle \rangle}{\langle \langle p_T \rangle \rangle} 
\left . \frac{d \bar p_T}{d\langle r \rangle} \right |_{\langle r \rangle = \langle \langle r \rangle \rangle}.
\label{sigsig2}
\end{eqnarray}
This result bears similarity to the formula derived by Ollitrault \cite {Ollitrault:1991xx}, 
where ${\sigma_{\rm dyn}}/{\langle \langle p_T \rangle \rangle} \sim \sigma(\langle s\rangle)/\langle \langle s \rangle \rangle$, with $s$ denoting the entropy density.
The derivative in Eq.~(\ref{sigsig2}) 
can be computed numerically without difficulty by running two simulations at each centrality.
We do it by comparing the average momenta obtained for the squeezed and stretched cases, as described above.
Then
\begin{eqnarray}
\frac{\sigma_{\rm dyn}}{\langle \langle p_T \rangle \rangle}=
\frac{\langle \langle p_T \rangle \rangle_- -\langle \langle p_T \rangle \rangle_+}
{\langle \langle p_T \rangle \rangle_- +\langle \langle p_T \rangle \rangle_+}, \label{dyn}
\end{eqnarray} 
where $-$ and $+$ indicate the squeezed and stretched cases.

\begin{figure}[b]
\begin{center}
\includegraphics[angle=0,width=0.47 \textwidth]{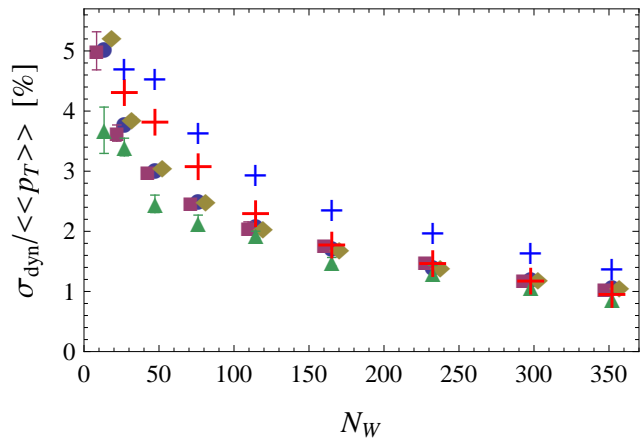} 
\end{center}
\vspace{-5mm}
\caption{(Color online) Comparison of the theoretical predictions for the scaled dynamical fluctuations, 
$\sigma_{\rm dyn}/\langle \langle p_T \rangle \rangle$ (for $\sqrt{s_{NN}}=200$~GeV), to the experimental data from the STAR collaboration
\cite{Adams:2005ka}. The lower (upper) crosses indicate our results for the wounded nucleon model (mixed model). The experimental data
range from $\sqrt{s_{NN}}=20$~GeV (triangles), through 130~GeV (squares), 62~GeV (diamonds), to $200$~GeV (dots). \label{fig:data}} 
\end{figure}

Our final result is shown in Fig.~\ref{fig:data}, where we compare the theoretical points to the experimental 
data from the STAR collaboration \cite{Adams:2005ka}. 
We note a strikingly good agreement between our calculation and the experiment, in particular for the 
standard wounded nucleon model. The mixed model, which is more realistic than the 
wounded-nucleon model, overshoots the data by about 20\%, producing even more fluctuations than 
needed. This may suggest that the coefficient 
${d \bar p_T}/{d\langle r \rangle} |_{\langle r \rangle = \langle \langle r \rangle \rangle}$ in Eq. (\ref{sigsig}) 
is somewhat too large. The value 
of this coefficient incorporates all the dynamics (the initial condition, hydrodynamics, statistical 
hadronization). Modifying these ingredients, not to mention the inclusion of viscosity effects, etc., may 
modify the value. Nevertheless, we note a proper magnitude of the effect and the 
correct dependence on centrality, with an approximate scaling 
$\sigma_{\rm dyn}(\langle p_T \rangle)/\langle \langle p_T \rangle \rangle \sim{1}/\sqrt{N_w}$. Also, since the 
results of Fig.~\ref{fig:basic} very weakly depend on $\sigma_{NN}$, 
to the extent that the hydrodynamic ``pushing'' is similar at various energies, our results should weakly depend on the incident energy, 
which is a desired experimental feature. We remark that the described mechanism works 
independently of the charge of particles. In order to describe 
the charge dependence of fluctuations ({\em e.g.} as observed in Ref.~\cite{Adams:2003uw}) other mechanisms are necessary.   

In conclusion, one can straightforwardly reproduce bulk of the dynamical event-by-event transverse momentum fluctuations, as measured at 
RHIC, with the mechanism based on fluctuations of the initial size, which are then carried over by hydrodynamics to the 
fluctuations of the transverse flow velocity, and consequently to the transverse momentum of the produced particles. 
The hydrodynamic pushing is crucial in this scheme.
With a realistic hydrodynamics, which 
has been earlier used to uniformly describe the soft RHIC data, our analysis indicates that there may 
be little room for other, truly dynamical, sources 
of fluctuations, such as (mini)jets \cite{Adler:2003xq,Liu:2003jf}} or the formation of 
clusters at freezeout \cite{Broniowski:2005ae,Tomasik:2008fq}. 
Certainly, there are yet other sources of the $p_T$ correlations 
in addition to the afore-mentioned ones, such as the global momentum conservation, 
resonance decays, correlations from the elementary $NN$ collisions in the corona, however, these 
should be considered at the ``background'' mechanism of the size fluctuations described in this paper. 

WB is grateful to Jean-Yves Ollitrault for pointing out Ref.~\cite{Ollitrault:1991xx}.


\end{document}